\documentclass[12pt]{article}
\usepackage{lingmacros}
\usepackage{tree-dvips}
\usepackage{graphicx}
\usepackage{subcaption}
\usepackage{amsmath}
\usepackage{color}
\usepackage{amssymb}
\usepackage{authblk}

\newcommand{\fixme}[1]{{\color{black}{#1}}}
\title{Thermodynamics and Phase Transition of Spherically Symmetric Black Hole in de Sitter Space from R\'enyi Statistics}

\author[1]{Lunchakorn Tannukij \footnote{Email: l\_tannukij@hotmail.com}}
	\affil[1]{\small Theoretical and Computational Physics Group, Theoretical and Computational Science Center(TaCS),
Faculty of Science, King Mongkut’s University of Technology Thonburi, 126 Pracha Uthit Rd., Bang Mod, Thung Khru, Bangkok 10140, Thailand}

\author[2,3]{Pitayuth Wongjun \footnote{Email: pitbaa@gmail.com}}
	\affil[2]{The Institute for Fundamental Study, Naresuan University, 99 Moo 9, Tapho, Mueang, Phitsanulok 65000, Thailand}
	\affil[3]{Thailand Center of Excellence in Physics, Ministry of Higher Education, Science, Research and Innovation, 328 Si Ayutthaya Road, Bangkok 10400, Thailand}
		
\author[1,4]{Ekapong Hirunsirisawat \footnote{Email: ekapong.hir@mail.kmutt.ac.th}}
	\affil[4]{Learning Institute, King Mongkut’s University of Technology Thonburi, 126 Pracha Uthit Rd., Bang Mod, Thung Khru, Bangkok 10140, Thailand}

\author[1]{Tanapat Deesuwan \footnote{Email: tanapat.dee@kmutt.ac.th}}

\author[1]{Chatchai Promsiri \footnote{Email: chatchaipromsiri@gmail.com}}

\date{}

\begin{document}

\maketitle

\begin{abstract}
Schwarzschild black holes in a de Sitter background were studied in terms of their thermodynamics based on the R\'enyi statistics. This led to thermodynamically stable black hole configurations for some certain range of black hole radii; namely within this range the corresponding black holes have positive heat capacity. Moreover, for a certain background temperature there can exist at most three configurations of black hole; one among which is thermodynamically stable. These configurations were investigated in terms of their free energies, resulting in the moderate-sized stable black hole configuration being the most preferred configuration. Furthermore, a specific condition on the R\'enyi non-extensive parameter is required if a given hot spacetime were to evolve thermally into the moderate-sized stable black hole.
\end{abstract}

\section{Introduction and Motivations}
%- Our background is de Sitter
%
%-The BH in dS background is not thermodynamically stable.
%
%-On the other hand, AdS BH can be thermodynamically stable but not realistic.
%
%-When being treated in Renyi statistics, Sch-flat BH can be thermodynamically stable as if the Renyi statistics renders the Sch-flat BH to be more like in AdS background.

The cosmic accelerating expansion of our universe has been proven to exist through numerous observations \cite{Perlmutter:1997zf,Riess:1998cb,Eisenstein:2005su}. To model this expansion, theorists have come up with various theories, one of which is the cosmological constant model \cite{Einstein:1931ab}. The key feature of the cosmological constant model is that the cosmic expansion is driven by the constant, $\Lambda$, which is treated as an additional matter/energy spreading throughout our universe. This model has been known as one of the most successful, yet simplest, cosmological models because it fits nicely with observations \cite{Aghanim:2018eyx}.

With the presence of the cosmological constant, a mathematical spherically-symmetric black hole solution can be considered as a generalization of the standard Schwarzschild solution. Depending on the sign of $\Lambda$, the corresponding black hole solution can either represent a Schwarzschild black hole on a de Sitter background for positive value of $\Lambda$, formally known as the Schwarzschild-de Sitter (Sch-dS) solution, or that on an anti-de Sitter background for negative value of $\Lambda$, known as the Schwarzschild-anti-de Sitter (Sch-AdS) solution. Since our late-time universe is found observationally to correspond to positive cosmological constant \cite{Aghanim:2018eyx}, then if black holes were to exist in our universe, they should correspond to the Sch-dS black holes. Despite the fact that our universe prefers the positive cosmological constant, the Sch-dS has been proven to be thermodynamically unstable according to the conventional thermodynamical statistics \cite{Teitelboim:2002cv}.

If a black hole is treated as a thermodynamic system, the black hole can be considered to have its own temperature due to the presence of the corresponding Hawking radiation \cite{Hawking:1975iha}. Taking the Schwarzschild black hole as an example, the Hawking radiation produced around the event horizon of the black hole happens in such a way that the corresponding Hawking temperature decreases as it gains energy, indicating that the Schwarzschild black hole is a thermal object with negative heat capacity \cite{Hawking:1975iha} (see also \cite{Czinner:2015eyk}). In particular, for a Schwarzschild black hole with its temperature above the background temperature, it will evaporate away as it radiates while the temperature increases, whereas for that with temperature below the background temperature, it will keep increasing in its size while the temperature decreases. As a consequence, the black hole will never be in thermodynamic equilibriumwith the background. For the Sch-dS black hole, the situation is even worse; the presence of the positive cosmological constant makes the heat capacity even more negative, given that in this work all of the thermodynamic quantities are evaluated at the black hole's horizon and not at the cosmological horizon. %This situation is unfortunate for our universe with positive cosmological constant. 
On the other hand, although it appears to go against with observations, the Sch-AdS black hole exhibits very nice thermal behaviours; there exist configurations with positive heat capacity which makes the black hole a thermodynamically stable object \cite{Hawking:1982dh}. In other words, the Sch-AdS black holes can be in thermal equilibrium with the background in a stable way. These features of the Sch-AdS are very interesting theoretically and they have been studied in a variety of ways, especially in the application to address thermal behaviours of gauge theories through AdS/CFT correspondence.

Those previous predictions are based on one common idea which is that all of the thermodynamic behaviours are described using the Gibbs-Boltzmann statistics. In particular, the so-called Bekenstein-Hawking entropy of a black hole of interest follows the standard thermodynamics based on the Gibbs-Boltzmann statistics.
%, the so-called Bekenstein-Hawking entropy, which is stated to be proportional to the area of the black hole. 
However, the entropy being proportional to the area, usually known as the area law, suggests that a black hole is not an extensive system, i.e. its entropy is not proportional to its volume. This fact 
%suggests researchers to try studying the thermodynamics of black holes using different thermal statistics. 
leads to various studies on thermodynamics of black holes using more general statistics which applies to non-extensive systems.
One of simple approaches is to treat the Bekenstein-Hawking entropy as the Tsallis entropy \cite{Tsallis:1987eu}. However, the Tsallis entropy appears to have a difficulty in defining thermodynamic temperature because of its incompatibility with the zeroth law of thermodynamics \cite{Biro}. To fix such a problem, an appropriate formulation of entropy which utilizes the formal logarithm of the Tsallis entropy is formally known as the R\'enyi entropy \cite{Renyi}. R\'enyi entropy exhibits very interesting features in the black hole thermodynamics as it effectively modifies the thermodynamics as if the black hole is in the AdS background without the necessity of introducing the negative cosmological constant. For example, there is an interesting study on a Schwarzschild black hole in an asymtotically flat spacetime using the R\'enyi entropy \cite{Czinner:2015eyk}. Although, in the Gibbs-Boltzmann context the black hole is said to have a negative heat capacity, the R\'enyi statistics suggests that the Schwarzschild black hole can possibly have a positive heat capacity withing a certain range of parameters, which resembles the thermal property of the Sch-AdS black hole according to Gibbs-Boltzmann statistics. In addition, the R\'enyi statistic treatment has been proven to be very interesting in the context of a Reissner-Nordstrom black hole in which the thermodynamical stability of the hole exists and in this language the Reissner-Nordstrom black hole exhibits  intriguing thermal behaviours as if it is a Van der Waal liquid-gas system \cite{Promsiri}.

%Since according to the observational data, the simplest black hole in our universe should be explained well by the Sch-dS solution. 
As motivated by the observational data, since it could represent a black hole in our late-time universe, the Sch-dS black hole is considered in the present work. 
Although the Sch-dS black holes are thermodynamically unstable in the context of the Gibbs-Boltzmann statistics, it is interesting to study the thermal behaviours of the Sch-dS black holes using R\'enyi statistics. In Sec. \ref{BHindSsec}, the black hole thermodynamics according to R\'enyi statistics are investigated and it turns out that a Sch-dS black hole can be thermodynamically stable; namely it can have a positive heat capacity for some certain parameter setup.  Additionally, for an appropriate background temperature, there are up to three possible states of the Sch-dS black hole which can be in thermal equilibrium to the background and only one of them is thermally stable. The free energies corresponding to the R\'enyi statistics are also considered in Sec. \ref{FEsec} in order to find the preferred state among the three possible ones. Interestingly, it turns out that the one with positive heat capacity corresponds to the lowest free energy which renders it a most preferred state. Finally, we also investigate the situation where a given hot space can actually evolve into a Sch-dS black hole with positive heat capacity through the free energy analysis and we report the condition of the parameter setup for such a situation. We finish this article by giving concluding remarks in Sec. \ref{Conclusionsec}.

\section{Black Hole in de-Sitter Space}\label{BHindSsec}

A spherically symmetric black hole solution to the Einstein's field equation in the presence of the positive cosmological constant $\Lambda$, frequently known as the Schwarzschild-de Sitter (Sch-dS) solution, is given by the following line element,
\begin{eqnarray}
d s^{2}=-f(r) d t^{2}+\frac{1}{f(r)} d r^{2}+r^{2} d \Omega^{2},
\end{eqnarray}
where
\begin{eqnarray}
f(r)=1-\frac{2 M}{r}-\frac{\Lambda r^{2}}{3},
\end{eqnarray}
given that $M$ is an integration constant which corresponds to a mass of the black hole. At the event horizon defined as $r_{+}$, we have $f\left(r_{+}\right)=0$ and we can find the black hole mass as follows,
\begin{eqnarray}
M=\frac{r_{+}}{2}\left(1-\frac{\Lambda r_{+}^{2}}{3}\right). \label{BHmass}
\end{eqnarray}
Note that $M$ is positive when $r^2_+<\frac{3}{\Lambda}$. 
Interestingly, the Hawking temperature of the black hole becomes lowered in the de Sitter space.  This can be expressed through the formula $T=\frac{f^{\prime}\left(r_{+}\right)}{4 \pi}$. Then, we obtain

\begin{align} 
	T &= \frac{1}{4 \pi}\left(\frac{2 M}{r_{+}^{2}}- \frac{2\Lambda r_{+}}{3}\right), \\ 
	   &=  \frac{1}{4 \pi}\left(\frac{1}{r_{+}}-\frac{\Lambda r_{+}}{3}-\frac{2 \Lambda r_{+}}{3}\right), \\ 
    	&=  \frac{1}{4 \pi r_{+}}\left(1-\Lambda r_{+}^{2}\right). 
\end{align}

The Bekenstein-Hawking entropy, which follows the area law, is in the following form,
\begin{eqnarray}
S_{\text{BH}}=\frac{A}{4}=\pi r_{+}^{2}.
\end{eqnarray}
\fixme{From the entropy formula, the area law states that the entropy is proportional to the area of the thermal system. This is one of the main reasons as to why black holes have been argued to be non-extensive systems, which does not coincide with the usual thermodynamic concepts. As previously mentioned in the introduction, in order to understand such systems, there are several kinds of thermal statistics which incorporates non-extensivity into thermal systems. To this end, the Tsallis entropy is one of the possible candidates.}
%A black hole has been argued to be a non-extensive system. Its entropy can be expressed in several ways to include non-extensivity.  The Tsallis entropy is one of the candidate. 
However, using Tsallis entropy leads to the problem of the incompatibility with the zeroth law of thermodynamics \cite{Biro} (see also in Appendix \ref{AppA}).  There are a suggestion to mathematically transform it to R\'enyi entropy.  
%Writing Tsallis entropy as Bekenstein-Hawking entropy, 
\fixme{By assuming that the black hole entropy follows the Tsallis statistics,} the R\'enyi entropy is expressed in the following form \cite{Renyi},
\begin{eqnarray}
S_{\text{R}} &=& \frac{1}{\lambda} \ln \left(1+\lambda S_{\text{BH}}\right) \nonumber \\ 
&=& \frac{1}{\lambda} \ln \left(1+\lambda \pi r_{+}^{2}\right)  \label{RenyiEntropy}
\end{eqnarray}
where $-\infty<\lambda<1$ is the non-extensive parameter. Note that in the limit $\lambda\to0$, this formula reduces to the Bekenstein-Hawking entropy $S_{BH}$. 

\begin{figure}[h!]
\begin{center}
\includegraphics{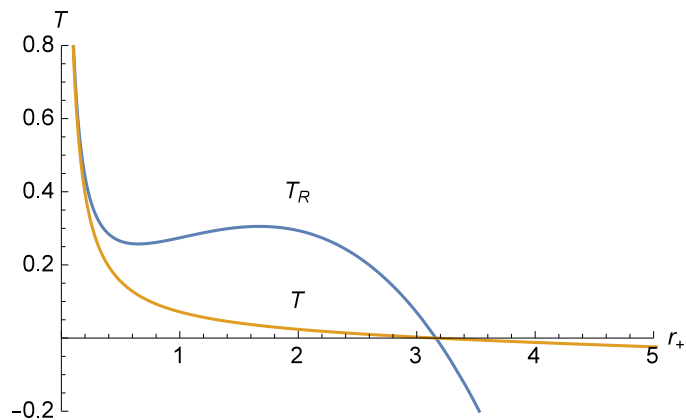}
\end{center}
\caption{The temperature profile corresponding to that of the R\'enyi entropy (blue) and the Bekenstein-Hawking temperature (yellow) of the Sch-dS black hole are shown here. The corresponding parameters are set as follows, $\lambda=0.9,\Lambda=0.02.$}\label{TRplotfig}
\end{figure}

Using R\'enyi statistics, the black hole stability is investigated through its temperature profile.  The R\'enyi temperature can be determined through $T_\text{R}=\frac{dM}{dS_{\text{R}}}$.  Let us take a few steps before that.   Using Eq.~\eqref{BHmass}, the differentiation of the black hole  mass is
\begin{eqnarray}
d M=\frac{1}{2}\left(1-\Lambda r_{+}^{2}\right) d r_{+}. \label{dM}
\end{eqnarray}
It is obvious that the extremum of the mass profile is at $r^2_+=\frac{1}{\Lambda}$ and since the second derivative of $M$ is always negative for positive values of $r_+$, this extremum is a maximum. By knowing that $M=0$ when $r_+=0$, the mass $M$ monotonically increases in $r_+$ for $0<r^2_+<\frac{1}{\Lambda}$.
By using Eq.~\eqref{RenyiEntropy}, the differentiation of the R\'enyi entropy is
\begin{eqnarray}
d S_{\text{R}} &= & \frac{1}{\lambda} \frac{2 \lambda \pi r_{+}}{\left(1+\lambda \pi r_{+}^{2}\right)} d r_{+}, \nonumber \\
            &= & \frac{2 \pi r_{+}}{1+\lambda \pi r_{+}^{2}} d r_{+}. \label{dS}
\end{eqnarray}
Using Eq.~\eqref{dM} and Eq.~\eqref{dS}, we obtain the R\'enyi temperature $T_\text{R}$ in the form of the Hawking temperature $T$ as follows
\begin{eqnarray}
T_{\text{R}} &=& \frac{1}{4 \pi r_{+}}\left(1-\Lambda r_{+}^{2}\right)\left(1+\lambda \pi r_{+}^{2}\right) \nonumber \\ &=&\left(1+\lambda \pi r_{+}^{2}\right) T. \label{TR}
\end{eqnarray}
Note that the black hole of size as parametrically large as $r^2_+=\frac{1}{\Lambda}$ corresponds to a zero temperature, which is related to the so-called Nariai limit \cite{Nariai1,Nariai2}.
The above expression can be rewritten in the following form,
\begin{eqnarray}
T_{\text{R}}=\frac{1}{4 \pi r_{+}}\left(1+(\lambda \pi-\Lambda) r_{+}^{2}-\Lambda \lambda \pi r_{+}^{4}\right). \label{TR2}
\end{eqnarray}
The R\'enyi temperature profile is expressed in Fig.\,\,\ref{TRplotfig}.
The R\'enyi temperature of the Sch-dS black hole has a very interesting feature. Particularly, the non-extensive parameter in the R\'enyi statistics provides the thermal effect as if the system is in a background with negative cosmological constant \cite{Czinner:2015eyk}. As a result, the Schwarzschild black hole thermodynamics according to the R\'enyi statistics is similar to the Bekenstein-Hawking thermodynamics of a Schwarzschild-anti-de Sitter black hole \cite{Czinner:2015eyk}. In this case where the Sch-dS black hole is considered, the non-extensive parameter changes the temperature profile of the Sch-dS black hole in an interesting way. In particular, the R\'enyi temperature is modified from the Bekenstein-Hawking temperature such that there exist two extrema; one minimum and one maximum. To this end, let us consider the following,
%Then, we would like to discuss about black hole stability through the plot $T_{\text{R}}$ versus $r_{+}$.  Because of this, we consider
\begin{eqnarray}
d T_{\text{R}} &=&  \left(-\frac{1}{4 \pi r_{+}^{2}}+\frac{\lambda \pi-\Lambda}{4 \pi}-\frac{3 \Lambda \lambda }{4} r_{+}^2 \right) d r_{+} \nonumber \\ 
&=&-\frac{1}{4 \pi r_{+}^{2}}\left(1-(\lambda \pi-\lambda) r_{+}^{2}+3 \Lambda \lambda \pi r_{+}^{4}\right) d r_{+}. 
\end{eqnarray}
By considering $\frac{dT_\text{R}}{dr_{+}}=0$, we know that $T_\text{R}$ becomes extremal at $r_{+}$ satisfying 
\begin{eqnarray}
3 \Lambda \lambda \pi r_{+}^{4}-(\lambda \pi-\Lambda) r_{+}^{2}+1=0\label{extremalcond}
\end{eqnarray}
The roots of this equation follows
\begin{eqnarray}
r_{+1,2}^2=\frac{\lambda \pi-\Lambda \pm \sqrt{(\lambda \pi-\Lambda)^{2}-12 \Lambda \lambda \pi}}{6 \Lambda \lambda \pi}\label{rminmax}
\end{eqnarray}
{From Eq.\,\,\eqref{rminmax}, to have both roots to be positive, we need $\lambda\pi-\Lambda>0$ from which, according to the definition of $\Lambda>0$ corresponding to the de Sitter background, it can be implied that $\lambda>0$.  Moreover, we also need $\left(\lambda\pi-\Lambda\right)^2>12\Lambda\lambda\pi$ so that the roots are not complex numbers. 
By requiring $\left(\lambda\pi-\Lambda\right)^2-12\Lambda\lambda\pi>0$, first we recall that 
\begin{align}
\left(\lambda\pi-\Lambda\right)^2-12\Lambda\lambda\pi=\left(\lambda\pi-\Lambda\left(7+\sqrt{48}\right)\right)\left(\lambda\pi-\Lambda\left(7-\sqrt{48}\right)\right), \label{factorlamb}
\end{align}
meaning that for $\left(\lambda\pi-\Lambda\right)^2-12\Lambda\lambda\pi>0$ we require either $\lambda\pi>\Lambda\left(7+\sqrt{48}\right) \approx 13.93\Lambda$ or $\lambda\pi<\Lambda\left(7-\sqrt{48}\right) \approx 0.07180\Lambda$. Since the latter case means that $\lambda\pi<\Lambda$, then the only applicable case is that $\lambda\pi>\Lambda\left(7+\sqrt{48}\right)$. This marks the lower bound for the value of $\lambda$.

Since in this scenario there is a local minimum in temperature, it can be proven that the minimum temperature is always positive for any allowed set of parameters, i.e. it cannot be as low as zero. Since at the minimum temperature, it also satisfies Eq.\,\,\eqref{extremalcond}. Then we can find the minimum temperature from simplifying Eq.\,\,\eqref{TR} using Eq.\,\,\eqref{extremalcond} as follows,
\begin{align}
T_{\text{R},min}&=\frac{1}{6\pi r_{+min}}\left(2+\left(\lambda\pi-\Lambda\right)r^2_{+min}\right),\nonumber
\\
&= \frac{1}{6\pi r_{+min}}\left(2+\frac{\lambda\pi-\Lambda}{6\Lambda\lambda\pi}\left[\left(\lambda\pi-\Lambda\right)-\sqrt{\left(\lambda\pi-\Lambda\right)^2-12\Lambda\lambda\pi}\right]\right).
\end{align}
Since from requiring that $\lambda\pi-\Lambda>0$ and $\Lambda>0$, the square bracket in the second line is always positive, rendering the minimum temperature always positive.
}

The corresponding heat capacity can be evaluated as
\begin{eqnarray}
C=\frac{d M}{d T_\text{R}}=\frac{-2 \pi r_{+}^{2}\left(1-\Lambda r_{+}^{2}\right)}{\left(1-(\lambda \pi-\Lambda) r_{+}^{2}+3 \Lambda \lambda \pi r_{+}^{4}\right)}.\label{Ceqn}
\end{eqnarray}

\begin{figure}[h!]
\begin{center}
\includegraphics[scale=1]{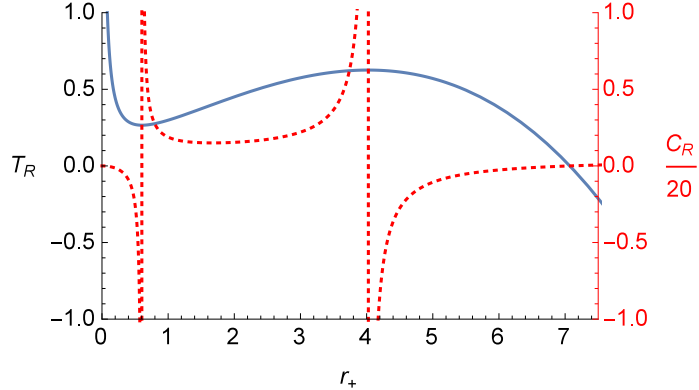}
\end{center}
\caption{The heat capacities corresponding to black holes of different radii (red-dotted line) is shown in comparison with the temperature profile (blue line). The corresponding parameters are set as follows, $\lambda=0.9,\Lambda=0.02.$}\label{CRplotfig}
\end{figure}

Let us consider the case where both of the $r_{+1,2}^2$ are positive, in other words the case which coincides with Fig.\,\,\ref{TRplotfig}. 
Note that the denominator in Eq.\,\,\eqref{Ceqn} is exactly the same function as in Eq.\,\,\eqref{extremalcond} whose roots are those in Eq.\,\,\eqref{rminmax}. Since $\lambda>0,\Lambda>0$ (which renders the denominator a convex parabola in an argument $r^2_+$), it can then be implied that the denominator in  Eq.\,\,\eqref{Ceqn} is negative when $r_{+1}^2<r_+^2<r_{+2}^2$. For the numerator in Eq.\,\,\eqref{Ceqn}, it can be shown that both the roots; $r_{+1,2}^2$, are less than $\frac{1}{\Lambda}$ by the following. From the requirement that $\lambda\pi-\Lambda>0$, we can express the roots as follows,
\begin{align}
r_{+1,2}^2&=\frac{\lambda \pi\left(1-\frac{\Lambda}{\lambda\pi}\right) \pm \lambda\pi\sqrt{1-\frac{14\Lambda}{\lambda\pi}+\frac{\Lambda^2}{\left(\lambda\pi\right)^2}}}{6 \Lambda \lambda \pi},\nonumber
\\
%&\approx \frac{1}{6\Lambda}\left[\left(1-\frac{\Lambda}{\lambda\pi}\right)\pm\left(1+\frac{1}{2}\left(-\frac{14\Lambda}{\lambda\pi}+\frac{\Lambda^2}{\left(\lambda\pi\right)^2}\right)\right)\right],\nonumber
%\\
%&=\frac{1}{6\Lambda}\left[\left(1-\frac{\Lambda}{\lambda\pi}\right)\pm\left(1-\frac{7\Lambda}{\lambda\pi}+\frac{\Lambda^2}{2\left(\lambda\pi\right)^2}\right)\right],\nonumber
%\\
%&=\frac{1}{6\Lambda}\left(2-\frac{8\Lambda}{\lambda\pi}\right), \frac{1}{6\Lambda}\left(\frac{6\Lambda}{\lambda\pi}-\frac{\Lambda^2}{2\left(\lambda\pi\right)^2}\right),\nonumber
%\\
&\approx\frac{1}{3\Lambda}\left(1-\frac{4\Lambda}{\lambda\pi}+\mathcal{O}(2)\right), \frac{1}{\lambda\pi}\left(1-\frac{\Lambda}{12\left(\lambda\pi\right)}+\mathcal{O}(2)\right).\label{approxrminmax}
\end{align}
{From Eq.\,\,\eqref{approxrminmax}, it is obvious that as $\lambda\pi-\Lambda>0$, $r_{+1}^2<\frac{1}{3\Lambda}<\frac{1}{\Lambda}$ and $r_{+2}^2<\frac{1}{\lambda\pi}<\frac{1}{\Lambda}$. For a Sch-dS black hole of a size corresponding to $r_{+1}^2<r_+^2<r_{+2}^2$, the numerator in Eq.\,\,\eqref{Ceqn} is thus always negative. Note that these bounds are in consistency with the positive black-hole mass bound, $\Lambda r^2_+\leq3$. Taking these bounds into account, we can conclude that for the range of interest; $r_{+1}^2<r_+^2<r_{+2}^2$, the corresponding heat capacity is always positive, meaning that the holes in this range are all locally thermodynamically stable whereas for those with sizes smaller than $r^2_{+1}$ or larger than $r^2_{+2}$. The corresponding heat capacities for various sizes of black holes are shown in Fig.\,\,\ref{CRplotfig}.}

\begin{figure}[h!]
\begin{center}
\includegraphics{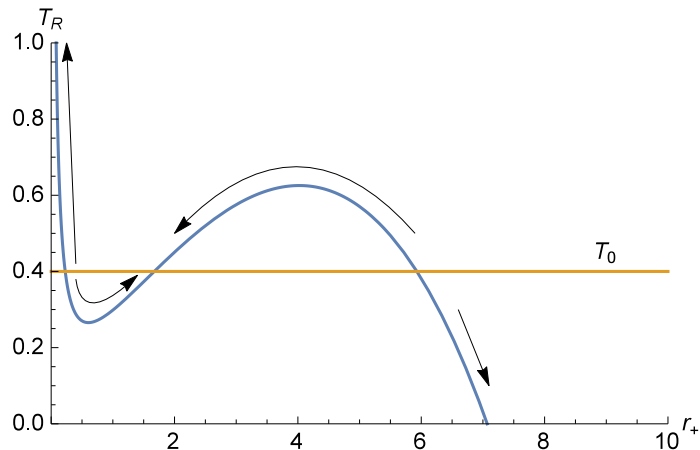}
\end{center}
\caption{The temperature profile corresponding to that of the R\'enyi entropy is shown here with the constant temperature $T_0$ indicating that there are 3 possible black holes. The arrows denote directions of evolution for each kind of black holes when surrounded by a system of temperature $T_0$. The corresponding parameters are set as follows, $\lambda=0.9,\Lambda=0.02, T_0=0.4.$}\label{TRplotcomparefig}
\end{figure}

{The R\'enyi temperature profile with two-extrema nature also exhibits very interesting thermal behaviours. For example, if a system is in a temperature which lies in between the local minimum temperature and the local maximum temperature, there can exist 3 kinds of black holes (the constant temperature line cuts through 3 points in the temperature profile as in Fig.\,\,\ref{TRplotcomparefig}). As previously mentioned that the mass $M$ monotonically increases in $r_+$, then the $r_+$ axis can be also treated as an increasing $M$ axis. The unstable small holes of higher temperature will evaporate away, while those of lower temperature will evolve into stable moderate-size black holes as they receive thermal radiation from the system and gain mass, resulting in an increase in $r_+$. The unstable large black holes of higher temperature will shrink as they lose masses through their own thermal radiation and eventually evolve into the stable moderate-size black holes. On the other hand, those of lower temperature will keep receiving thermal radiation from the system, and finally evolve into the zero-temperature black holes which are the largest possible holes. Note that these behaviours arise because the black holes in consideration have negative heat capacities.}

%\begin{figure}[h!]
%\begin{center}
%\includegraphics{TRplotphasetrans}
%\end{center}
%\caption{The temperature profile corresponding to that of the R\'enyi entropy is shown here in the case where the two extrema coincide, leading to a phase transition in the parameter space. The parameter setup is the following, $\lambda=0.9,\Lambda=0.203001$}\label{TRplotphasefig}
%\end{figure}

%{Remarks from P'Nguang : 
%Since the two extrema coincide with one another when the discriminant, $\left(\lambda\pi-\Lambda\right)^2-12\Lambda\lambda\pi$, vanishes, this could mark the point in a parameter space where a phase transition happens, as shown as an example in Fig.\,\,(\ref{manyphasetransfig}). In particular, we can show this phase transition behaviour through a $r_+,\lambda$ phase space as in Fig.\,\,\ref{phasespacefig}.}

\begin{figure}[h!]
\begin{center}
\includegraphics{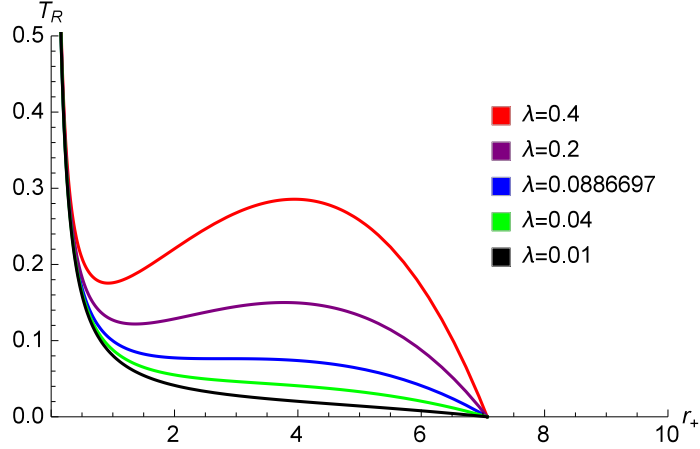}
\end{center}
\caption{The various R\'enyi temperature profiles corresponding to various $\lambda$ as $\Lambda=0.02$ are shown here. %The parameter setup is the following, $\lambda=0.4,0.2,0.0886697,0.04,0.01$ corresponding to the graphs of red, purple, blue, green, and black respectively. 
The blue graph corresponds to a parameter set which exhibits the phase transition between a phase of no possible thermally-stable black hole into that of possible ones.}\label{manyphasetransfig}
\end{figure}

\begin{figure}[h!]
\begin{center}
\includegraphics{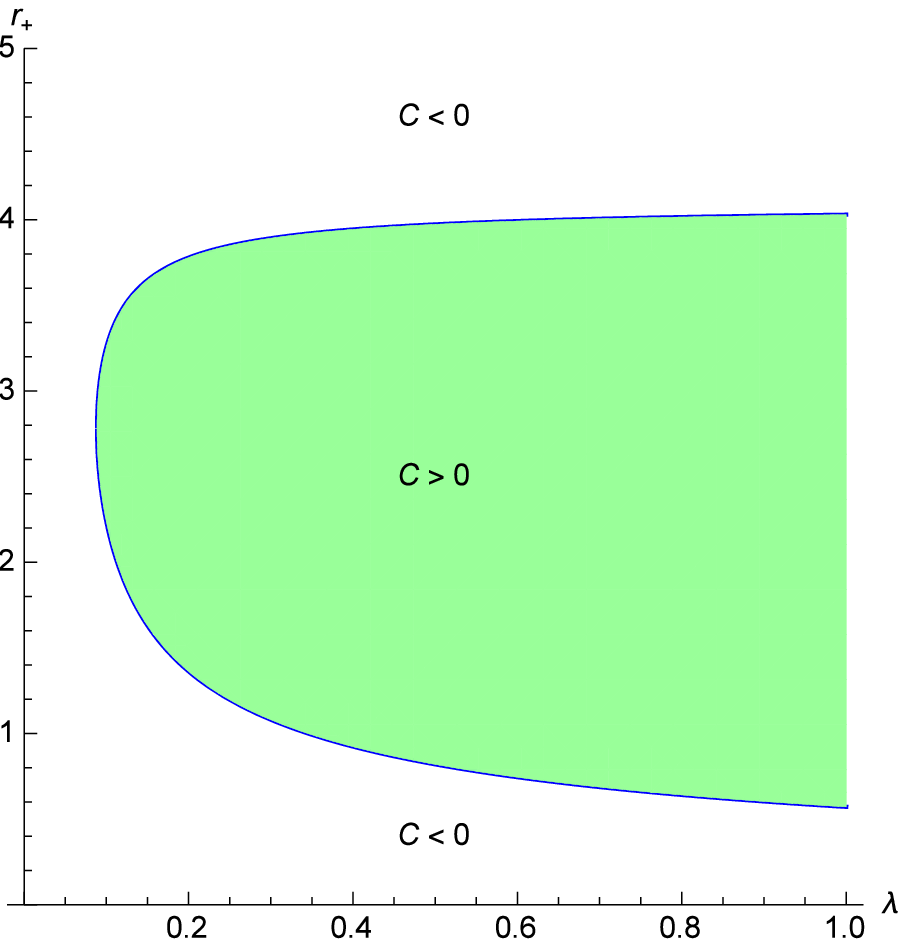}
\end{center}
\caption{The various configurations of $(\lambda,r_+)$ are shown here given that $\Lambda=0.02$. The lines denote boundaries between phases of different behaviours of the heat capacities, thus different thermodynamical stabilities, for each configuration of $(\lambda,r_+)$. The green-shaded region denotes the configurations corresponding to positive heat capacity while the unshaded region corresponds to negative heat capacity configurations.}\label{phasespacefig}
\end{figure}

The mentioned features arise because of the effect of the non-extensive parameter, $\lambda$. This gives rise to the two extrema in the R\'enyi temperature profile. There also exists a specific parameter setup where the two extrema merge and become only one extremum; particularly it satisfies the vanishing discriminant, or $\left(\lambda\pi-\Lambda\right)^2-12\Lambda\lambda\pi=0$. In this case, $r^2_{+1}=r^2_{+2}$ and the thermodynamically stable black hole does not exist because the corresponding heat capacity is always negative. This specific set of parameters can be viewed as a point of phase transition in a parameter space, namely it defines a boundary in the parameter space separating the region in which there exists a locally thermodinamically stable black hole and the region of no possible thermodinamically stable black hole. This phase transition is shown graphically in Fig.\,\,\ref{manyphasetransfig} where various parameter setups are shown in comparison. %Moreover, the existence of thermally-stable black holes are illustrated in Fig.\,\,\ref{phasespacefig} according to various values of $\lambda$.
Moreover, the allowed region corresponding to the existence of thermodynamically-stable black holes can be illustrated in a $\lambda-r_+$ phase space as shown in Fig.\,\,\ref{phasespacefig}.

\fixme{From Fig.\,\,\ref{phasespacefig}, it is obvious that for a given parameter setup, there is a minimum value of $\lambda$ below which no thermodynamically stable black hole can be found. The minimum $\lambda$ can be found analytically through Eq.\,\,\eqref{factorlamb}.}
{%This phase transition can be found analytically through Eq.\,\,\eqref{factorlamb}. 
Since we require $\lambda\pi>\Lambda$ for the temperature to have the extrema on the positive side of $r_+$, then the condition for the phase transition point is that $\lambda\pi=\Lambda\left(7+\sqrt{48}\right)$ for a particular value of $\Lambda$.}

\section{Free energies}\label{FEsec}

According to the unique feature of the R\'enyi temperature of the Sch-dS black hole, if the background temperature lies in between the minimum and the maximum temperature, then there are at most three configurations of black hole; one of which is thermodynamically stable where the other two are not. This could lead one to think that there are three possible states of black hole which can be formed from hot gas in the spacetime of given temperature. These three states can be shown that they are not equally likely to happen by considering each of their corresponding free energies. 
{By formulating thermodynamic quantities following the R\'enyi statistics, we can define the corresponding Helmholtz free energy as a Legendre transformation of the mass,
\begin{align}
F_R&=M-T_R S_R,
\\
&=\frac{r_+}{2}\left(1-\frac{\Lambda r^2_+}{3}\right)-\frac{\left(1+\left(\lambda\pi-\Lambda\right)r^2_+-\Lambda\lambda\pi r^4_+\right)}{4\pi\lambda r_+}\ln\left(1+\lambda\pi r^2_+\right). \label{freeenergy}
\end{align}
\begin{figure}[h!]
\begin{center}
\includegraphics{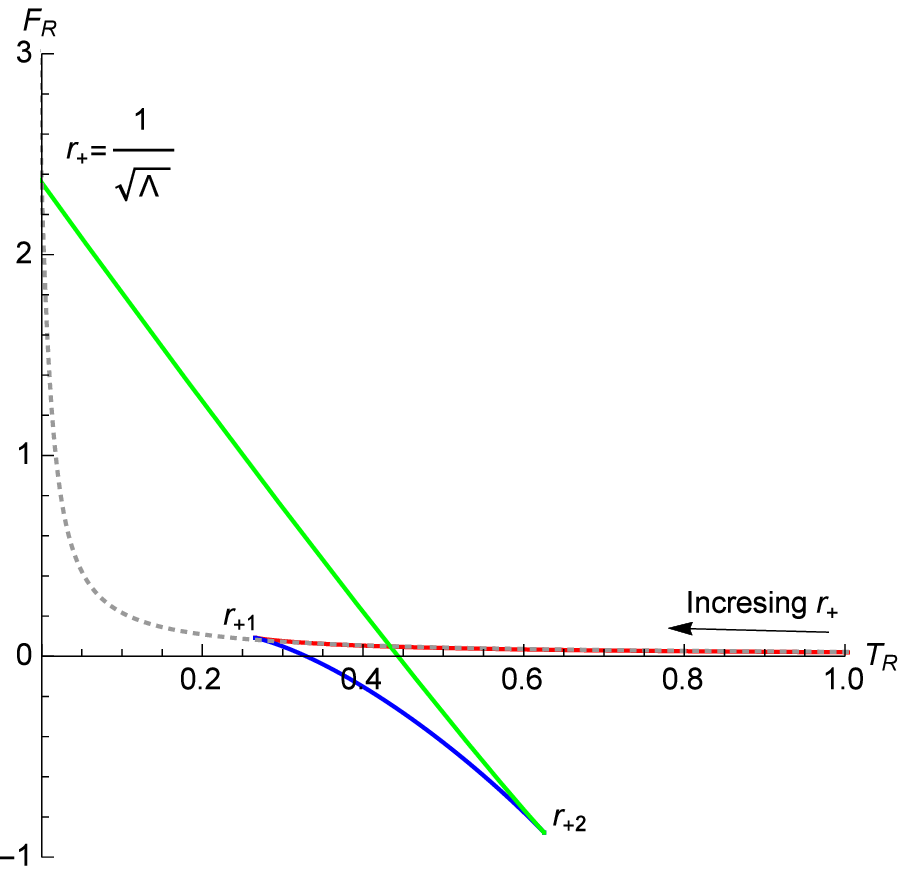}
\end{center}
\caption{The free energy profile according to each temperature is shown here. The parameter setup is the following, $\lambda=0.9,\Lambda=0.02$. The graph shows the various values of free energy according to different possible configurations for a given temperature; the red line, blue line, and green line correspond to configurations in which $0<r_+<r_{+1},\,\,r_{+1}<r_+<r_{+2},\,\, \text{and }r_{+2}<r_+<\frac{1}{\sqrt{\Lambda}}$ respectively. The dotted-grey line indicates the free energy profile when $\lambda=0,\Lambda=0.02$ according to which the R\'enyi free energy reduces into the Gibbs-Boltzmann free energy.}\label{FRTRparaplot}
\end{figure}

A plot of the free energy against the temperature is plotted in Fig.\,\,\ref{FRTRparaplot}. It can be seen that for a given temperature the free energy corresponding to the positive-heat-capacity black hole is the lowest, indicating that the thermodinamically stable black hole is preferred. \fixme{This behaviour arises from the fact that the slope of the $F_R-T_R$ graph is negative R\'enyi entropy, specifically $\frac{dF_R}{dT_R}=-S_R$. Since the entropy is an increasing function in $r_+$ then as $r_+$ increases, the entropy increases which also makes the slope more negative. If we trace the $F_R-T_R$ graph from $r_+=0$ to $r_+=r_{+1}$ (the red line in Fig.\,\,\ref{FRTRparaplot}), we will see that the graph between this range of $r_+$ results in increasing $F_R$ but decreasing $T_R$ and the slope will be more negative as $r_+$ is increasing. However, when we trace the graph from $r_+=r_{+1}$ to $r_+=r_{+2}$ (the blue line in Fig.\,\,\ref{FRTRparaplot}), the graph is such that $F_R$ decreases as $T_R$ increases but the slope still keeps being negatively steeper. Lastly, as we trace the graph from $r_+=r_{+2}$ to $r_+=\frac{1}{\sqrt{\Lambda}}$ (the green line in Fig.\,\,\ref{FRTRparaplot}), $F_R$ begins to increase again as $T_R$ decreases and the slope is still negatively steeper. These behaviours result in the lowest values of free energy corresponding to the section of graph in the range between $r_+=r_{+1}$ to $r_+=r_{+2}$ and suggest that the black holes of sizes $r_{+1}<r_+<r_{+2}$, the thermally stable ones, correspond to the lowest free energy at a given temperature and is more likely to form than the other configurations.} 

Given that a spacetime of consideration is at some certain temperature, the free energy of the hot spacetime itself is zero, meaning that in order for a black hole to form itself from the hot spacetime, the free energy of the hole must be negative. This can be realized by considering the local minimum of the free energy (because the maximum and the minimum of possible $r_+$ result in the free energies not less than zero, meaning that the global extrema of the free energy cannot be global minima.). From $dF_R=-S_R dT_R$, we know that the local maximum of the temperature is automatically the local minimum of the free energy, making $r_{+2}$ corresponds to the local minimum of $F_R$ as shown in Fig.\,\,\ref{FRplotfig}. 
\begin{figure}[h!]
\begin{center}
\includegraphics[scale=1]{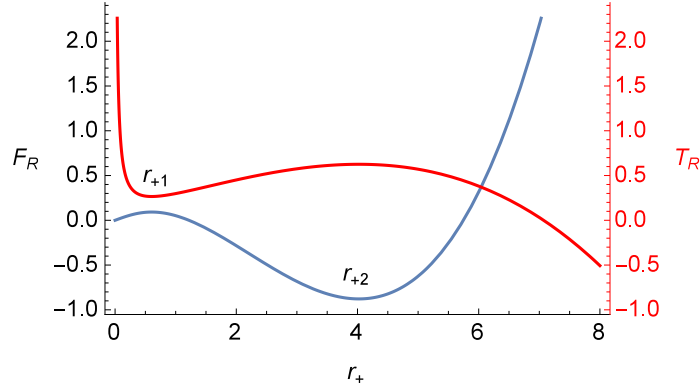}
\end{center}
\caption{The free energy profile according to each temperature is shown here. The parameter setup is the following, $\lambda=0.9,\Lambda=0.02$. The graph shows the various values of free energy according to different possible configurations for a  given temperature.}\label{FRplotfig}
\end{figure}
Since $r_{+2}$ satisfies Eq.\,\,\eqref{extremalcond} then we can use Eq.\,\,\eqref{extremalcond} to simplify Eq.\,\,\eqref{freeenergy} as follows,
\begin{align} 
F_R(r_{+2})=\frac{1+8\lambda\pi r^2_{+2}+\Lambda r^2_{+2}-\left(6+3\left(\lambda\pi-\Lambda\right)r^2_{+2}\right)\ln\left(1+\lambda\pi r^2_{+2}\right)}{18\lambda\pi r_{+2}}. \label{Fzero}
\end{align}
By requiring that $F_R<0$, then the numerator in Eq.\,\,\eqref{Fzero} must be negative. After substituting in the expression for $r^2_{+2}$, we have the following condition,
\begin{flalign}
&8-\epsilon-\epsilon^2+\left(\epsilon+8\right)\sqrt{1-14\epsilon+\epsilon^2}&\nonumber
\\
&<3\left(1+10\epsilon+\epsilon^2+\left(1-\epsilon\right)\sqrt{1-14\epsilon+\epsilon^2}\right)\ln\left(\frac{1+5\epsilon+\sqrt{1-14\epsilon+\epsilon^2}}{6\epsilon}\right)&, \label{Fzerocond}
\end{flalign}
where we have defined $\epsilon\equiv\frac{\Lambda}{\lambda\pi}$. In order to satisfy Eq.\,\,\eqref{Fzerocond}, it turns out numerically that we must have $\epsilon<0.0328$. This bound is even stronger than the previous bound, \textit{i.e.} $\epsilon\equiv\frac{\Lambda}{\lambda\pi}<\frac{1}{7+\sqrt{48}}\approx 0.0718$, required to have both minimum and maximum in the temperature profile. This suggests that for a given temperature, which lies between the minimum and the maximum temperature, the bound $\epsilon<0.0328$, must be additionally satisfied in order to have a thermodynamically stable black hole as a preferred state in a hot spacetime.

}

\section{Conclusions}\label{Conclusionsec}

In this article, we have studied thermodynamics of a Schwarzschild black hole in a de Sitter background in the approach of R\'enyi statistics. The Sch-dS black hole is known to have a negative heat capacity when its thermodynamic quantities are evaluated on the hole's horizon based on the Gibbs-Boltzmann statistics. The possible reason behind this may be that the Gibbs-Boltzmann statistics is not applicable to non-extensive systems such as black holes. One of the approaches for treating non-extensive systems is to consider thermodynamics based on R\'enyi statistics. When the black hole is considered in the language of R\'enyi statistics, compared with that in Gibbs-Boltzmann statistics, the corresponding temperature profile shows an increasing temperature for a specific range of black hole radii, indicating an existence of positive heat capacity which can be seen as an example in Fig.\,\,\ref{TRplotfig} and Fig.\,\,\ref{CRplotfig}. This interesting behaviour only appears when the non-extensive parameter $\lambda$ and the cosmological constant $\Lambda$ obey the following bound:  $\epsilon\equiv\frac{\Lambda}{\lambda\pi}<\frac{1}{7+\sqrt{48}}\approx 0.0718$.
%a suitable parameter setup is chosen such that $\lambda\pi > \Lambda\left(7+\sqrt{48}\right)\approx 13.93\Lambda$. 
For a certain value of background temperature, there exists a temperature such that it lies in between the local minimum and the local maximum temperatures, resulting in three possible configurations of thermal black hole as shown in Fig.\,\,\ref{TRplotcomparefig}. In particular, the configuration with the moderate black hole radius is thermodynamically stable while the other two, namely those with the smaller and bigger radii, are unstable. According to our findings, the R\'enyi statistics of the Sch-dS black hole suggests that the smaller holes can either evaporate away if the holes have higher temperature than that of the background or evolve into the moderate-sized stable black holes if they have lower temperature than that of the background. For the bigger holes, they can either evolve into the moderate-sized stable ones if their temperatures are higher than that of the background or evolve into zero-temperature black holes which are the largest possible holes given that black holes of negative temperature do not exist. 

From the free energy analysis, these three possible configurations correspond to different values of free energies, suggesting that these three configurations are preferred differently for a thermal background of some certain temperature. From the previous analysis on their free energies, the moderate-sized stable black hole which is in thermal equilibrium with the background appears to be the preferred configuration compared with the other two (unstable) configurations as shown in Fig.\,\,\ref{FRTRparaplot}. Furthermore, in order for a hot spacetime to evolve itself into a spacetime with a thermal Sch-dS black hole, the free energy of the preferred configuration of the black hole must be lower than that of the hot spacetime, which is zero. Requiring that the free energy of the preferred black hole is less than zero, a stronger bound of the parameter setup has been found to satisfy Eq.\,\,\eqref{Fzerocond} or numerically $\epsilon<0.0328$. This bound might suggest that if the real black holes were to be explained by R\'enyi statistics, then the R\'enyi non-extensive parameter, $\lambda$, would be possibly determined by the cosmological constant, $\Lambda$. 

According to the previously mentioned analyses, we found that based on R\'enyi statistics, it is possible to have thermodynamically stable black holes in dS background if non-extensivity is taken into account. Moreover, the non-extensive parameter $\lambda$ behaves as if it produces thermal effects usually seen on black holes in AdS background even though the actual background is not AdS. In particular, it can be seen from the second term in Eq. \eqref{TR2} that the non-extensive parameter $\lambda$ seems to compete with the cosmological constant $\Lambda$ since they are of opposite signs. Thanks to these feaures, this study may give us possibilities in describing black holes in our universe in the context of their thermodynamical stabilities or even broadening the research field towards AdS/CFT correspondence. 

\section*{Acknowledgements}

The authors acknowledge King Mongkut's University of Technology Thonburi for funding Postdoctoral Fellowship to LT. PW is supported by the Thailand Research Fund (TRF) through grant no. MRG6180003. PW also would like to thank the Theoretical and Computational Physics Group (TCP) for hospitality
during this work was in progress.

\appendix
\section*{Appendix}
\section{Zeroth Law Compatibility}\label{AppA}

In Sec. \ref{BHindSsec}, the R\'enyi entropy is considered as thermodynamic entropy of the black hole system instead of the Tsallis entropy. The reason behind this is that the Tsallis entropy which obey the following composition,
\begin{align}
S_{12}=S_1+S_2+\lambda S_1 S_2,
\end{align}
is not compatible with the zeroth law of thermodynamics which is crucial in defining temperature of a thermal system. In order to see this, let us consider a thermal isolated system with constant entropy, $S$. We can always consider this system as being composed of two weakly-interacting subsystems, each of which has entropy of $S_1$ and $S_2$. Moreover, let us assume that the entropy composition rule is that of Gibbs-Boltzmann statistics as follows,
\begin{align}
S=S_1+S_2.
\end{align}
If the subsystems are allowed to exchange only thermal energy to one another, since the whole system is isolated, then according to the conservation of energy, the thermal energy gained by the subsystem $1$ must be from the subsystem $2$,
\begin{align}
T_1 dS_1=-T_2 dS_2.
\end{align}
If the system enter a thermal equilibrium, $T_1=T_2$, then we obtain $dS_1+dS_2=d\left(S_1+S_2\right)=dS=0$ which implies that there is no change in entropy which is reasonable for an isolate system. This statement is true only in the context of the Gibbs-Boltzmann entropy which obeys $S=S_1+S_2$. On the other hand, when the Tsallis entropy is considered instead, the above analysis will imply a change in total entropy even an isolated system is assumed, showing a contradiction between the Tsallis entropy and the zeroth law of thermodynamics. The more detailed investigation is reported in Ref. \cite{Biro}.

\end{document}